\documentclass[12pt]{article}
\pdfoutput=1
\usepackage{amsfonts,amsthm,amsmath,amssymb,graphicx,hyperref,youngtab}
\usepackage{cite}

\newcommand{\be}{\begin{eqnarray}}
\newcommand{\ee}{\end{eqnarray}}
\newcommand{\nn}{\nonumber}

\renewcommand{\t}{\tilde}

\begin{document}

\begin{center}

\vspace{0.9 cm}

{\LARGE \bf{Non-Linear Resonance in Relativistic Preheating\\\vspace{0.2cm}}}

\vspace{1.1 cm} {\large Bret Underwood\footnote{{\ttfamily {bret.underwood@plu.edu }}} and Yunxiao Zhai\footnote{{\ttfamily {zhaiya@plu.edu}}}}

\vspace{0.9 cm}
\vspace{0.2 cm}{{\it Department of Physics, Pacific Lutheran University, Tacoma, WA 98447}}

\thispagestyle{empty}
\vspace{1.5cm}

{\bf Abstract}
\end{center}

\begin{quotation}

Inflation in the early Universe can be followed by a brief period of preheating, resulting in rapid and non-equilibrium particle
production through the dynamics of parametric resonance.
However, the parametric resonance effect is very sensitive to the linearity of the reheating sector.
Additional self-interactions in the reheating sector, such as non-canonical kinetic terms like the DBI Lagrangian,
may enhance or frustrate the parametric resonance effect of preheating.
In the case of a DBI reheating sector, preheating is described by parametric resonance of
a damped relativistic harmonic oscillator. 
In this paper, we illustrate how the non-linear terms in the relativistic oscillator
shut down the parametric resonance effect.
This limits the effectiveness of preheating when there are non-linear self-interactions.

\end{quotation}

\setcounter{page}{0}
\setcounter{tocdepth}{2}

%%%%%%%%%%%%%%%%%%%%%%%%%%%%%%%%%%%%%%%
%%%%%%%%%%%%%%%%%%%%%%%%%%%%%%%%%%%%%%%
\section{Introduction}
%%%%%%%%%%%%%%%%%%%%%%%%%%%%%%%%%%%%%%%
%%%%%%%%%%%%%%%%%%%%%%%%%%%%%%%%%%%%%%%

The energy that drives inflation in the very early Universe is trapped in the homogeneous inflaton scalar field $\phi$.
As inflation ends and the inflaton field settles down into its minimum, this energy is transferred to observable matter and radiation
through an epoch of reheating.
When the inflaton couples to another scalar field $\chi$, called the ``reheaton" field, the dynamics often contain a period
of explosive particle production preceding the usual perturbative reheating called preheating (see \cite{InflationReview,ReheatReview,preheatingLong} 
for an introduction to preheating).

For example, the simple Lagrangian
\be
{\mathcal L} = \frac{1}{2} (\partial \phi)^2 - V(\phi) + \frac{1}{2} (\partial \chi)^2 - \frac{1}{2} g^2 \phi^2 \chi^2 - \frac{1}{2} m_{\chi}^2 \chi^2\,, \label{eq:CanonLagrangian}
\ee
describes the interaction between the inflaton and reheating fields. In an expanding Universe the equation of 
motion for large scale modes of $\chi$ becomes that of a damped harmonic oscillator with a time-dependent frequency:
\be
\ddot \chi(t) + 3 H \dot \chi(t) + \left(m_\chi^2 + g^2 \phi_0(t)^2\right) \chi(t) = 0\, , \label{eq:ChiEOMIntro}
\ee
where $\phi_0(t)$ is the time-dependent background solution for the inflaton field. At the end of inflation, the inflaton
oscillates about its potential minimum, resulting in an oscillating frequency for the reheaton field in (\ref{eq:ChiEOMIntro}).
A harmonic oscillator with a time-dependent frequency as in (\ref{eq:ChiEOMIntro}) can exhibit explosive growth in the
amplitude of $\chi$ through the non-linear phenomenon of parametric resonance (see \cite{LandauLifshitz,Nonlinear} for an introduction to parametric
resonance).

The interaction described in (\ref{eq:CanonLagrangian}-\ref{eq:ChiEOMIntro}) is a simple toy model for the physics of reheating and preheating --- in 
practice, the inflaton and reheaton sectors could be considerably more complicated.
In particular, additional interactions between and within the inflaton and reheaton sectors
in (\ref{eq:CanonLagrangian}-\ref{eq:ChiEOMIntro}) may enhance, or frustrate, the
parametric resonance effect of preheating.

One type of self-interactions, known as non-canonical kinetic terms, arise naturally in low-energy effective
field theories \cite{kinflation,Gelaton,Franche:2009gk,Karouby:2011xs}:
\be
{\mathcal L} = p(X_\phi,\phi) + \t p(X_\chi, \chi) - \frac{1}{2} g^2 \phi^2 \chi^2\,, \label{eq:AllNCLagrangian}
\ee
where $X_\phi \equiv -\frac{1}{2} (\partial \phi)^2$, $X_\chi \equiv -\frac{1}{2} (\partial \chi)^2$.
Note that we have left the interaction between the inflaton and reheaton sectors unchanged; some previous work which
has explored modified couplings between these sectors can be found in \cite{Lachapelle:2008sy}.

There has been considerable interest in the use of non-canonical kinetic terms such as those in (\ref{eq:AllNCLagrangian})
to describe alternative models of inflation \cite{kinflation,DBI,DBISky}. Non-canonical kinetic terms are interesting because perturbations about a cosmological
background can have a sound speed $c_s^2$ different than $1$, and can have interesting observational signatures such as violation of
the single-field consistency relation and enhanced non-gaussianity \cite{kinflationPert,DBI,DBISky,NonGauss}.
Recent cosmological data suggests that the sound speed of the inflaton during inflation cannot be too small, $c_s \geq 0.02$
\cite{PlanckOverview,PlanckCosmology,PlanckInflation,PlanckIsotropy,PlanckNonGauss}.
%Thus, during inflation the inflaton field behaves effectively like a canonical kinetic term,
%\be
%p(X_\phi, \phi) \approx \frac{1}{2} (\partial \phi)^2\, .
%\ee
In this paper, we are interested in investigating the role that non-canonical kinetic terms in the reheating sector alone can play in
the dynamics of preheating, so
%Similar constraints, however, are not present for the reheaton field. Taking this point of view, 
we will consider the inflaton-reheaton
Lagrangian to consist of a canonical inflaton field coupled to a non-canonical reheaton field:
\be
{\mathcal L} =  \frac{1}{2} (\partial \phi)^2 - V(\phi) + \t p(X_\chi, \chi) - \frac{1}{2} g^2 \phi^2 \chi^2\,. \label{eq:ReheatNCLagrangian}
\ee
We have chosen to couple the inflaton and reheaton sectors with a simple quartic cross-coupling, which has the
advantage that it does not lead to tachyonic effective masses for $\chi$.
In principle, other cross-couplings between the sectors such as $\lambda \phi \chi^2$ could be considered, as well as higher-order powers
of $\phi$ and $\chi$ such as
%\be
${\mathcal L}_{int} \sim - g_{(n,m)} \phi^n \chi^m/\Lambda^{4-n-m}\,.$
%\ee
A general analysis of all such possible terms involves many non-linear terms of different types, and as such is beyond the scope of this initial analysis.

As discussed in \cite{Gelaton,Franche:2009gk}, non-canonical kinetic terms of the form $\t p(X_\chi, \chi)$ can arise as an effective field theory below 
some scale of new physics $\Lambda$. Integrating out the physics above $\Lambda$ leads to an effective field theory in terms
of some non-renormalizable operators
\be
\t p(X_\chi,\chi) = \frac{1}{2} (\partial \chi)^2 - \frac{1}{2} m_\chi^2 \chi^2 + \sum_{n>4} c_n \frac{{\mathcal O}_n}{\Lambda^{n-4}}\,.
\ee
To be more precise, we will consider $\t p(X_\chi, \chi)$ to be of the Dirac-Born-Infeld (DBI) form \cite{DBI,DBISky}:
\be
\t p(X_\chi,\chi) = -\Lambda^4 \left[\sqrt{1-2 X_\chi/\Lambda^4}-1\right] - \frac{1}{2} m_\chi^2 \chi^2\,, \label{DBI}
\ee
where we will take the ``warp factor" $\Lambda^4$ to be a constant for simplicity.
This Lagrangian can arise as a low-energy effective field theory for the motion of a D-brane in warped
extra dimensions in string theory.

The equation of motion for long-wavelength modes of $\chi$ in an expanding background arising from (\ref{eq:ReheatNCLagrangian},\ref{DBI})
is then:
\be
\frac{\ddot \chi}{(1-\frac{\dot \chi^2}{\Lambda^4})^{3/2}} + 3 H \dot \chi + (m_\chi^2 + g^2 \phi(t)^2)\chi = 0\,. \label{RelMathieuIntro}
\ee
Now the reason for choosing the specific DBI Lagrangian is clear --- this equation of motion
has the form of a damped relativistic harmonic oscillator with a time-dependent effective frequency, so we can apply much of our
intuition from that simple model to understand preheating with non-canonical kinetic terms.
This is no accident --- the DBI action (\ref{DBI}) is a generalization of the relativistic point particle action to an spacetime filling
dynamical membrane, where the field $\chi$ denotes the embedding of the membrane in an extra dimension.
We will take (\ref{RelMathieuIntro}) as our starting point to understand how non-linearities introduced
by non-canonical kinetic terms affect the dynamics of preheating.
As will be discussed later, a more general Lagrangian with non-linear terms of the form 
${\mathcal L}_{non-linear} \sim c_8 {\mathcal O}(\chi^4)/\Lambda^4$ will exhibit similar behavior in preheating, 
so our conclusions are not just restricted to the DBI/relativistic oscillator case.

The role of the DBI Lagrangian (\ref{DBI}) in preheating has been considered before. In \cite{DBIPreheat1,DBIPreheat2,Zhang:2013asa}, 
the leading order effects of an inflaton
field with a DBI kinetic term for preheating were studied at low speeds. Ref \cite{Karouby:2011xs}, in contrast, considered the effects of a DBI inflaton
on preheating at high speeds. More recently, \cite{Child:2013ria} studied numerical lattice simulations of preheating for
an inflaton with a DBI kinetic term. All of these previous works consider the non-canonical kinetic term to be in the inflaton sector;
however, as discussed above, it is also possible for the reheaton sector to have non-canonical kinetic terms\footnote{
See also \cite{Chemissany:2011nq}, which studied the effects on preheating of non-linear terms in the reheaton sector arising
from a Generalized Uncertainty Principle.}.

Preheating is a complex phenomenon, involving not just parametric resonance but also rescattering and backreaction effects, and
inhomogeneous evolution. We do not intend to include all of these effects here; that approach is best suited to complex numerical
simulations, such as \cite{Felder:2000hq,Frolov:2008hy,Child:2013ria}. Instead, our goal is to attempt to gain some analytic and conceptual understanding of
how the non-linearities inherent in (\ref{RelMathieuIntro})
affect the onset of parametric resonance for preheating. As we will see below, analytic techniques will help us to generate some insight
into solutions of (\ref{RelMathieuIntro}), and shed light on the effect of non-linear terms on parametric resonance.

The rest of the paper is organized as follows. In Section \ref{sec:RelOscillator}, we describe the equation of motion for long-wavelength modes
of a DBI reheaton field, and cast it as a relativistic oscillator with a time-dependent frequency.
In Section \ref{sec:Properties}, we review some of the properties of the relativistic oscillator, including its dependence of
oscillation frequency with amplitude, and its behavior when subject to a resonant driving force.
In Section \ref{sec:RelResonance}, we analyze parametric resonance of the relativistic oscillator, making use of the technique
of multiple scales. We find there that in the traditional ``resonance band" of the non-relativistic oscillator where solutions
grow exponentially, solutions to the relativistic oscillator do not exhibit parametric resonance and their amplitudes are bounded.
In Section \ref{sec:DampedRel}, we perform a similar analysis for the damped relativistic oscillator, finding similar behavior.
In Section \ref{sec:Preheating} we discuss the implications of our analysis for preheating in the early Universe.
We conclude in Section \ref{sec:Conclusion} with some final remarks.

%%%%%%%%%%%%%%%%%%%%%%%%%%%%%%%%%%%%%%%
%%%%%%%%%%%%%%%%%%%%%%%%%%%%%%%%%%%%%%%
\section{The Relativistic Oscillator}
\label{sec:RelOscillator}
%%%%%%%%%%%%%%%%%%%%%%%%%%%%%%%%%%%%%%%
%%%%%%%%%%%%%%%%%%%%%%%%%%%%%%%%%%%%%%%

Our starting point is the Lagrangian (\ref{eq:ReheatNCLagrangian},\ref{DBI}) in a flat FRW spacetime
\be
ds^2 = -dt^2 + a(t)^2 d\vec{x}^2\,,
\ee
which gives rise to the long-wavelength equations of motion:
\be
&&H^2 = \left(\frac{\dot a}{a}\right)^2 = \frac{1}{3 M_p^2} \left[\frac{1}{2} \dot \phi_0^2 + V(\phi_0)+ \frac{1}{2} g^2 \phi_0^2 \chi^2
	+ \Lambda^4\left(\frac{1}{\sqrt{1-\dot\chi^2/\Lambda^4}}-1\right)+\frac{1}{2} m_\chi^2 \chi^2\right];\hspace{.3in} \label{HEOM}\\
&& \ddot \phi_0 + 3 H \dot \phi_0 + V'(\phi_0) = 0; \\
&& \frac{\ddot \chi}{(1-\dot \chi^2/\Lambda^4)^{3/2}} + 3 H \dot \chi + \left(m_\chi^2 + g^2 \phi_0(t)^2\right)\chi = 0\, .\label{eq:ChiEOM1}
\ee
For simplicity, we will take the inflaton to have the potential $V(\phi) = \frac{1}{2} m_\phi^2 \phi^2$, arising for example 
from a quadratic expansion of the inflaton potential about the inflaton's minimum.
For this choice of potential, at the end of inflation the inflaton oscillates about its minimum as a function of time as \cite{preheatingLong}:
\be
\phi_0(t) = \Phi(t) \sin(m_\phi t) = \frac{\Phi_0 \sin(m_\phi t)}{t}\, . \label{InflatonOscillations}
\ee
Inserting (\ref{InflatonOscillations}) into (\ref{eq:ChiEOM1}), we obtain the equation of motion for the long-wavelength
reheaton field during preheating:
\be
\frac{\ddot \chi}{(1-\dot \chi^2/\Lambda^4)^{3/2}} + 3 H \dot \chi + \left(m_\chi^2 + g^2 \Phi(t) \sin^2(m_\phi t)\right)\chi = 0\, .\label{eq:ChiEOM2}
\ee
By making the redefinition $\tau \equiv m_\phi t$, (\ref{eq:ChiEOM2}) can be written as:
\be
\frac{\chi''(\tau)}{(1-\chi'^2/v^2)^{3/2}} + \mu \chi'(\tau) + (A-2 q \cos( 2\tau)) \chi(\tau) = 0 \label{eq:ChiEOM3}\, ,
\ee
where ${}' \equiv \frac{d}{d\tau}$, $v \equiv \Lambda^2/m_\phi$ is the ``speed limit" of $\chi(\tau)$,
$\mu \equiv 3 H/m_\phi$ is a damping coefficient, and $A, q$ are defined as
\be
A \equiv \frac{m_\chi^2 + \frac{1}{2} g^2 \Phi^2}{m_\phi^2}, \hspace{.5in} q \equiv \frac{g^2 \Phi^2}{4 m_\phi^2}\, . \label{MathieuParameters}
\ee
Since the inflaton amplitude $\Phi(t)$ is decaying with time, in principle the parameters $A,q,\mu$ are also time-dependent.
However, for our analysis in Section 2 we will assume that these parameters are approximately constant so that their timescale
is much longer than the timescale for preheating. In Section 3 we will allow $A,q,\mu$ to be time-dependent, and find that this
does not affect our conclusions.

Equation (\ref{eq:ChiEOM3}) is the relativistic generalization of the damped Mathieu equation. At small speeds $|\chi'| \ll v$,
(\ref{eq:ChiEOM3}) reduces to the usual Mathieu equation. It is well known that the non-relativistic Mathieu
equation exhibits a phenomenon known as parametric resonance --- for certain values of $q, A$ the amplitude of $\chi(\tau)$
grows exponentially with time \cite{LandauLifshitz,Nonlinear}.
However, at high speeds the non-linearity inherent in (\ref{eq:ChiEOM3}) can be important.
In the rest of this section, we will focus on how the non-linearity of (\ref{eq:ChiEOM3}) affects the explosive growth
of parametric resonance. In the next section, we will apply these results to preheating.

%%%%%%%%%%%%%%%%%%%%%%%%%%%%%%%%%%%%%%%
\subsection{Properties of the Relativistic Oscillator}
\label{sec:Properties}
%%%%%%%%%%%%%%%%%%%%%%%%%%%%%%%%%%%%%%%

\begin{figure}[t]
\centering\includegraphics[width=.5\textwidth]{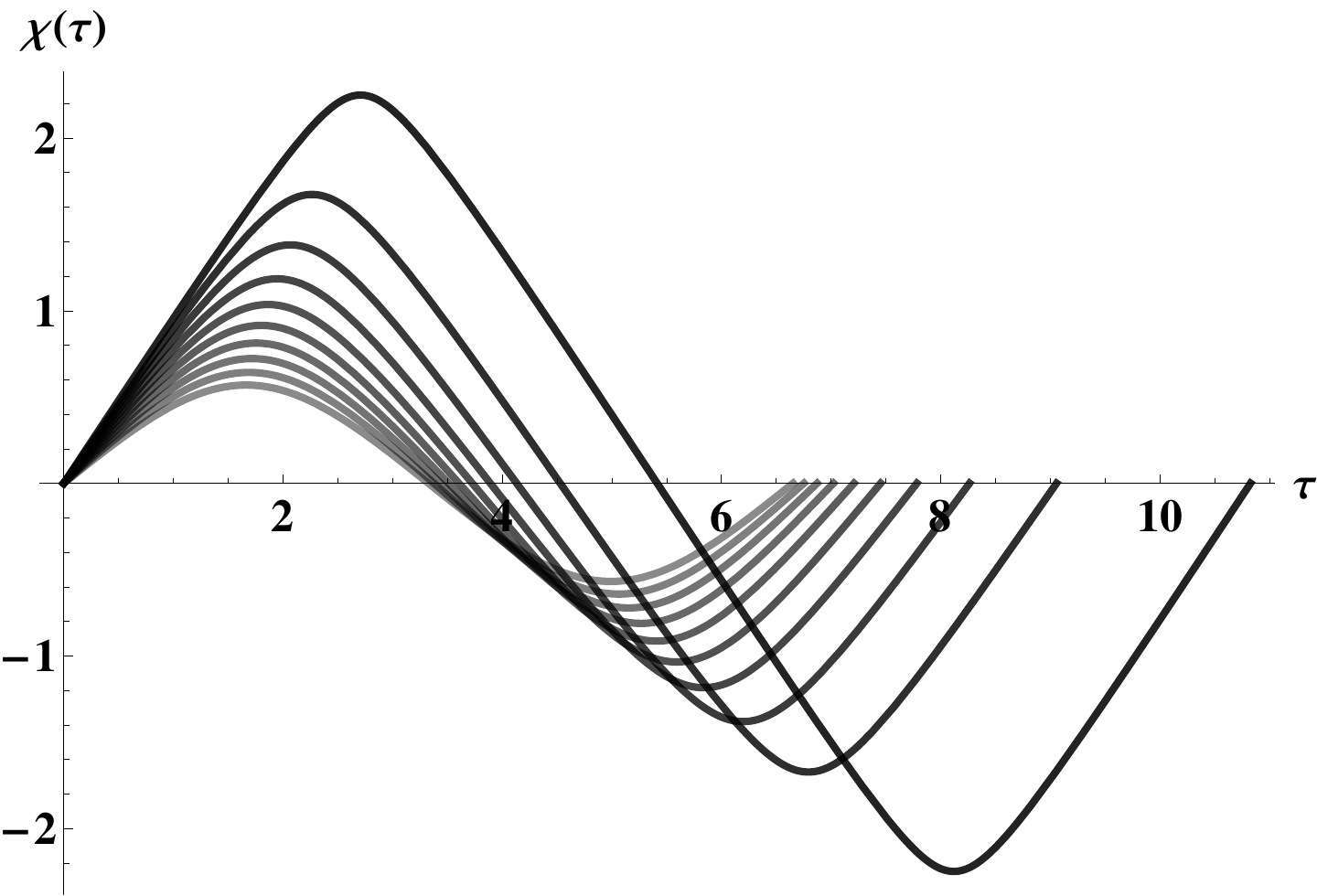}\includegraphics[width=.5\textwidth]{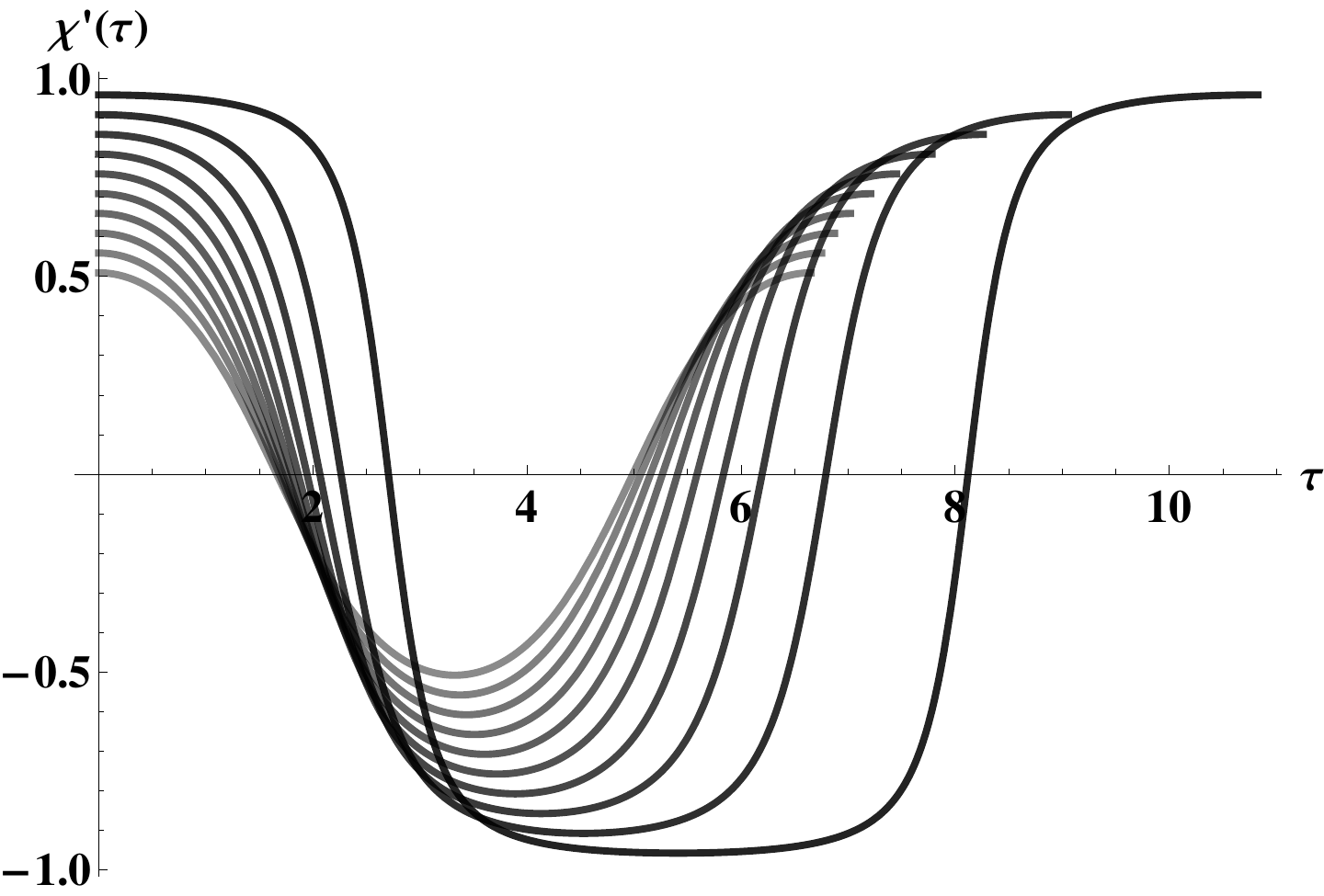}
\caption{The period of the relativistic harmonic oscillator (\ref{RelOscillator}) depends on the amplitude, in contrast to its
non-relativistic counterpart. This amplitude dependence arises because of the speed limit of the relativistic oscillator --- any increase in amplitude
must be accompanied by a decrease in frequency. This is shown in plots of $\chi$ versus $\tau$ (left) and $\chi'$ versus $\tau$ (right), where
we have set the relativistic speed $v = 1$.
}
\label{fig:RelativisticAmplitude}
\end{figure}
To begin, let us review the behavior of the relativistic harmonic oscillator in general, taking
$\mu = 0 = q$:
\be
\frac{\chi''(\tau)}{(1-\chi'^2/v^2)^{3/2}} + A \chi(\tau) = 0 \label{RelOscillator}\, .
\ee
It is possible to solve (\ref{RelOscillator}) in quadrature to obtain $\tau = \tau(\chi)$ in terms of elliptic integrals \cite{RelOscillator,Goldstein}
(see also \cite{RelDamped,RelDamped2,RelAnHarmonicOscillator} for more studies of the relativistic oscillator).
The details of these solutions will not be important to us here. However, there is one important feature to solutions
of (\ref{RelOscillator}) --- unlike the non-relativistic harmonic oscillator, the frequency of oscillation depends on the amplitude
of the oscillation.
The reason for this is quite simple --- as the amplitude increases, so does the speed $\chi'$ of the oscillator.
However, because of the speed limit imposed by (\ref{RelOscillator}) $|\chi'| < v$, when the oscillator reaches relativistic
speeds any further increase in amplitude must be accompanied by a corresponding decrease in the frequency of oscillation
so that the speed $|\chi'|$ does not exceed the speed limit. See Figure \ref{fig:RelativisticAmplitude} for a graphical illustration of this effect.
As we will discuss in more detail below, this has important consequences for the existence of resonance of the
relativistic oscillator.

\begin{figure}[t]
\centering\includegraphics[width=.8\textwidth]{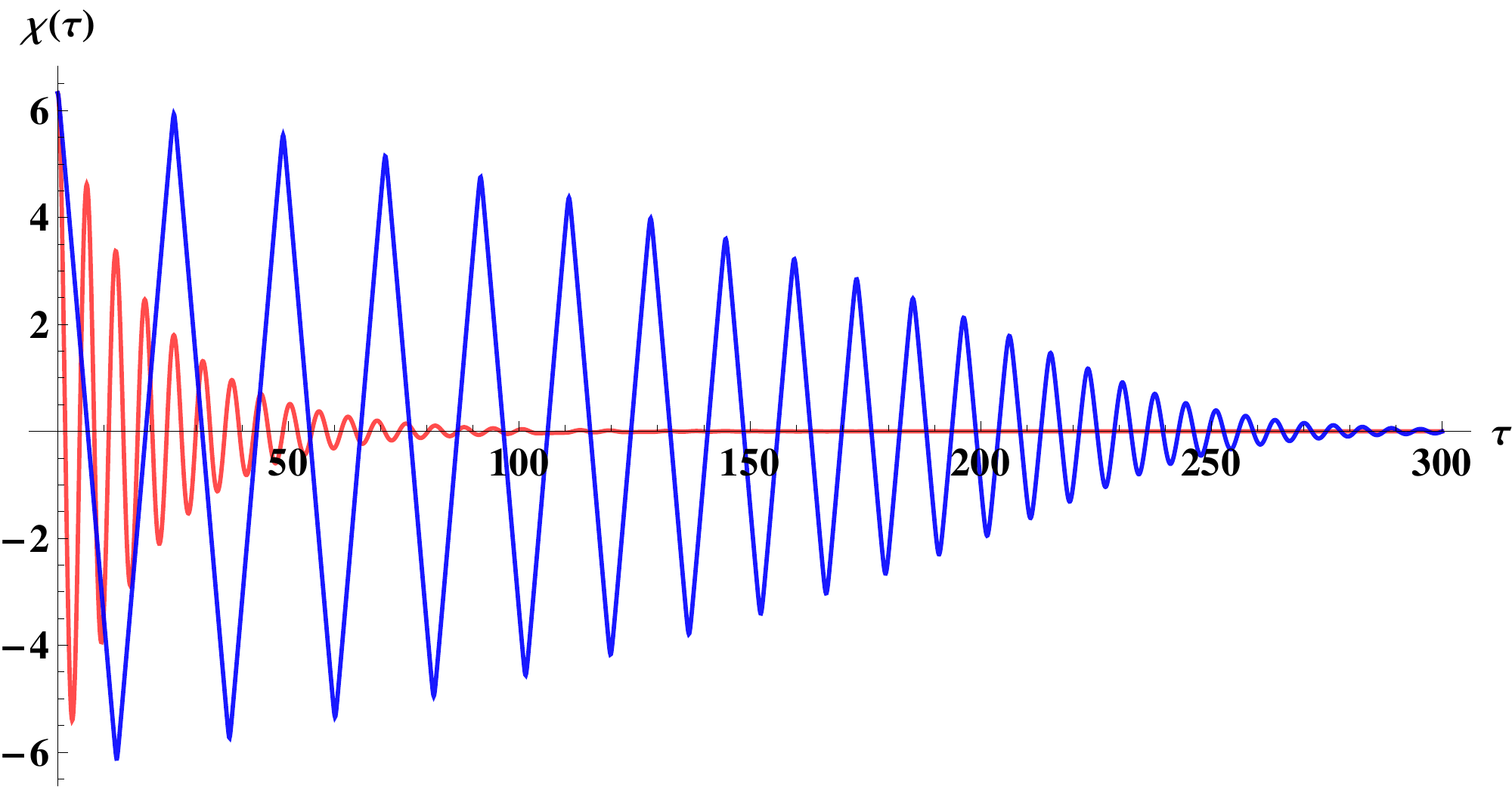}
\caption{The relativistic damped oscillator (\ref{DampedRelOscillator}) (blue, darker) ($v = 1$, $\mu = .1$, $A = 1$) loses energy less rapidly at large amplitudes than its non-relativistic counterpart 
(red, lighter) because the non-relativistic oscillator is able to reach higher speeds, and thus have higher damping energy losses.
}
\label{fig:RelativisticDamping}
\end{figure}

Including damping, (\ref{eq:ChiEOM3}) becomes the damped relativistic harmonic oscillator,
\be
\frac{\chi''(\tau)}{(1-\chi'^2/v^2)^{3/2}} + \mu \chi' + A \chi(\tau) = 0 \label{DampedRelOscillator}\, .
\ee
Solutions to (\ref{DampedRelOscillator}) behave much as in the non-relativistic case --- for small values of $\mu$, $\chi$
oscillates with a decaying amplitude (underdamped), while for large values of $\mu$, $\chi$ decays asymptotically to zero (overdamped).
The primary difference between solutions of the relativistic and non-relativistic damped oscillators is that the rate of decay is slower
for the former, as shown in Figure \ref{fig:RelativisticDamping}. The physical reason is straightforward to see --- since $\chi(\tau)$ 
is speed limited $|\chi'| < v$, at
high speeds the relativistic oscillator loses less energy to the velocity-dependent dissipation than its non-relativistic counterpart.

\begin{figure}[t]
\centering\includegraphics[width=\textwidth]{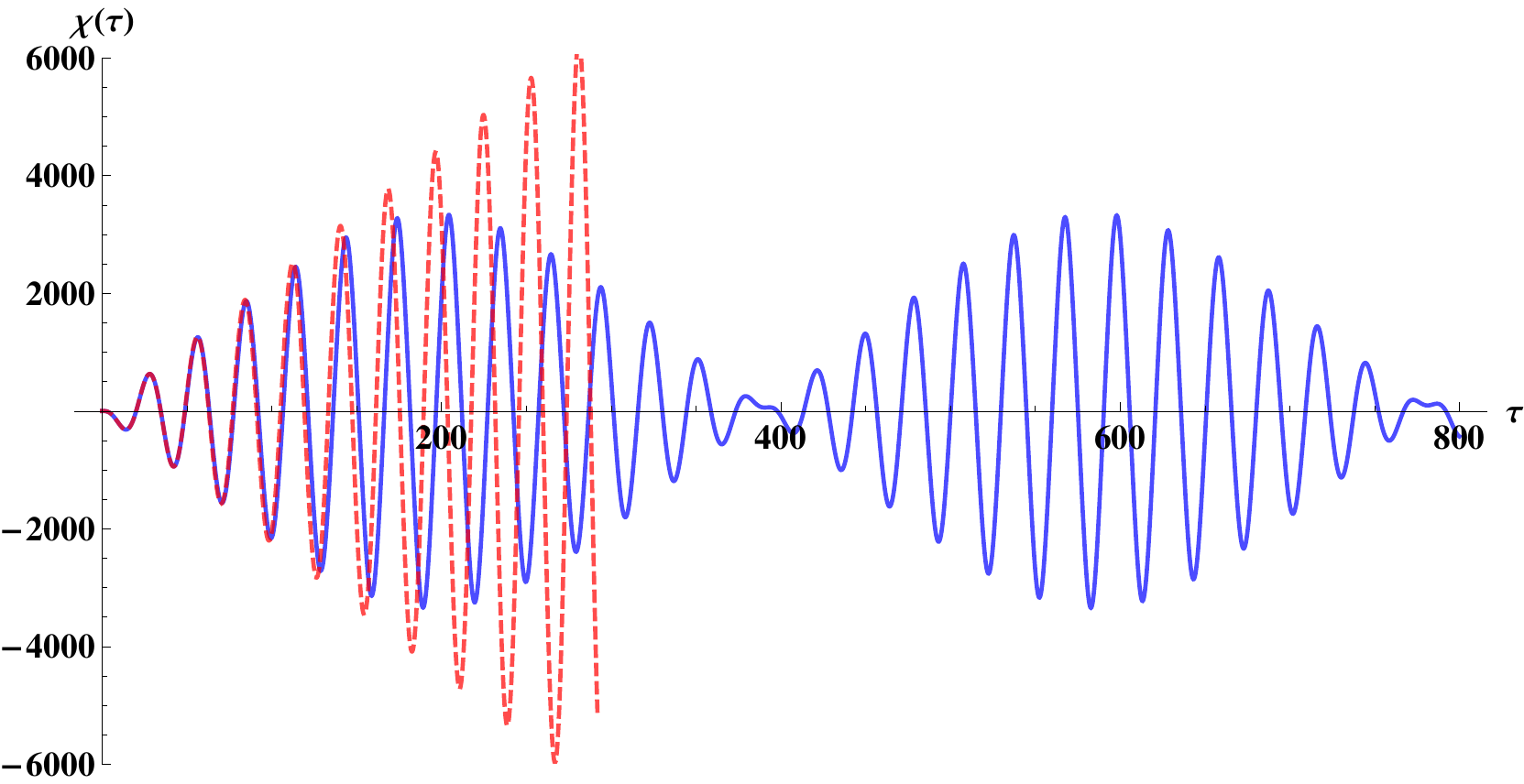}
\caption{The forced relativistic oscillator (blue, solid) ($v = 948$, $F_0 = 200\ A$) 
and forced non-relativistic oscillator (red, dashed) both initially grow in amplitude
when their natural frequency is equal to the driving frequency. However, once the amplitude is sufficiently large, the natural frequency of the
relativistic oscillator begins to decrease as in Figure \ref{fig:RelativisticAmplitude}, causing it to fall out of resonance. After the amplitude of the
relativistic oscillator has decreased sufficiently, the oscillator's natural frequency again matches the driving frequency, causing growth again.
This process repeats, resulting in a ``beating" like pattern of oscillation.
}
\label{fig:ForcedRelOscillator}
\end{figure}

Before studying the relativistic Mathieu equation (\ref{eq:ChiEOM3}) in more detail, let us consider one more similar system,
the forced relativistic harmonic oscillator:
\be
\frac{\chi''(\tau)}{(1-\chi'^2/v^2)^{3/2}} + A \chi(\tau) = F_0 \sin(\Omega t) \label{ForcedRelOscillator}\, .
\ee
As a reminder, in the forced non-relativistic harmonic oscillator, when the driving frequency is equal to the natural
frequency of the oscillator ($\Omega = A^{1/2}$), the system resonates and the amplitude
of $\chi(\tau)$ grows linearly with time. The behavior of (\ref{ForcedRelOscillator}) is much more complex, however.
For small initial amplitudes and speeds, where (\ref{ForcedRelOscillator}) behaves effectively like a non-relativistic system, 
$\chi$ will indeed begin to grow at the resonant frequency $\Omega = A^{1/2}$.
However as the amplitude grows the frequency of the relativistic oscillator $\chi$  also changes so that the oscillations
of $\chi$ are no longer in phase with the oscillations of the external force.
When this happens, instead of adding energy and increasing the amplitude, the external force begins to {\it remove}
energy and {\it decrease} the amplitude.
At some point the amplitude is again small enough that $\chi$ behaves like a non-relativistic oscillator, and thus the
frequency of oscillation again matches that of the external force and the amplitude begins to grow again.
This process repeats, resulting in a ``beating" pattern for $\chi(\tau)$, rather than unconstrained growth as in
the non-relativistic oscillator, as shown in Figure \ref{fig:ForcedRelOscillator}.
For large initial amplitudes, as discussed in \cite{RelChaos} the forced harmonic oscillator exhibits chaotic behavior rather than exponential growth.

%%%%%%%%%%%%%%%%%%%%%%%%%%%%%%%%%%%%%%%%%%%%%
\subsection{Relativistic Parametric Resonance}
\label{sec:RelResonance}
%%%%%%%%%%%%%%%%%%%%%%%%%%%%%%%%%%%%%%%%%%%%%

Finally, we are ready to analyze the relativistic Mathieu equation (\ref{eq:ChiEOM3}). For simplicity, let
us ignore damping for now and set $\mu = 0$ --- we will return to the effects of damping in the next subsection.
In general, solutions to (\ref{eq:ChiEOM3}) for arbitrary values of the parameters can only be obtained numerically,
and due to the non-linear nature of (\ref{eq:ChiEOM3}) their behavior can be difficult to understand.
Before we study these solutions, then, let us comment on some expected {\it qualitative} features
of solutions of (\ref{eq:ChiEOM3}).

The non-relativistic Mathieu equation has solutions which exhibit resonance (and thus unrestricted growth of the amplitude of $\chi$)
for small $q$ when the average natural frequency $A$ of the oscillator is approximately an integer $A \approx 1, 2, ...n$.
%the average natural frequency of the oscillator is approximately an integer multiple of the driving
%frequency divided by two, $A^2 \approx 1, 4, 9, ...n^2$.
From our knowledge of the driven relativistic oscillator discussed above, we know that while resonance may initially occur
for (\ref{eq:ChiEOM3}) when the initial amplitude is small, it will soon stop when the relativistic corrections to the frequency
cause the oscillations to be out of phase with the oscillations of the effective frequency $\omega^2_{eff} = A-2 q \cos 2\tau$.

Another perspective can be obtained by rearranging (\ref{eq:ChiEOM3}) in the absence of damping as
\be
\chi''(\tau) + (A-2 q \cos 2\tau) \left(1-\frac{\chi'^2}{v^2}\right)^{3/2} \chi(\tau) = 0\, . \label{RelMathieuRearranged}
\ee
As discussed before, as the amplitude of $\chi$ increases, so does the speed $|\chi'|$ until the speed limit is reached.
However, as the speed limit is approached the coefficient of the new effective frequency in (\ref{RelMathieuRearranged}),
namely $(A-2 q \cos 2\tau) \left(1-\frac{\chi'^2}{v^2}\right)^{3/2}$, vanishes so that at large speeds the oscillator becomes
a nearly free particle without resonance.
Thus, on basic principles we expect that while the non-relativistic Mathieu equation exhibits exponential growth and resonance
for certain values of $A, q$, the relativistic Mathieu equation (\ref{eq:ChiEOM3}) or (\ref{RelMathieuRearranged}) may contain
some initial growth, but should eventually saturate at some maximum amplitude.

Let us now check this expectation against solutions of (\ref{RelMathieuRearranged}). As mentioned before, it is difficult
to interpret generic numerical solutions of (\ref{RelMathieuRearranged}) for arbitrary values of the parameters.
However, some analytic progress can be made perturbatively when we set $q = \epsilon \hat q$, $v^2 = \hat v^2 \epsilon^{-1}$,
where $\hat q, \hat v \sim {\mathcal O}(1)$ and $\epsilon \ll 1$ is a small parameter, so that $q \ll 1$ and $v^2 \gg 1$.
It is then possible to expand (\ref{RelMathieuRearranged}) to leading order in $\epsilon$ as
\be
\chi'' + (A-2 \hat q \epsilon \cos 2\tau) \chi = \frac{3}{2} \epsilon A \frac{\chi'^2}{{\hat v}^2} \chi + {\mathcal O}(\epsilon^2)\, . \label{PertMathieu}
\ee
While (\ref{PertMathieu}) seems like a significant simplification of (\ref{RelMathieuRearranged}), as we will discuss solutions
of (\ref{PertMathieu}) 
contain all of the same qualitative features as numerical solutions to (\ref{RelMathieuRearranged}).

\begin{figure}[t]
\centering \includegraphics[width=.5\textwidth]{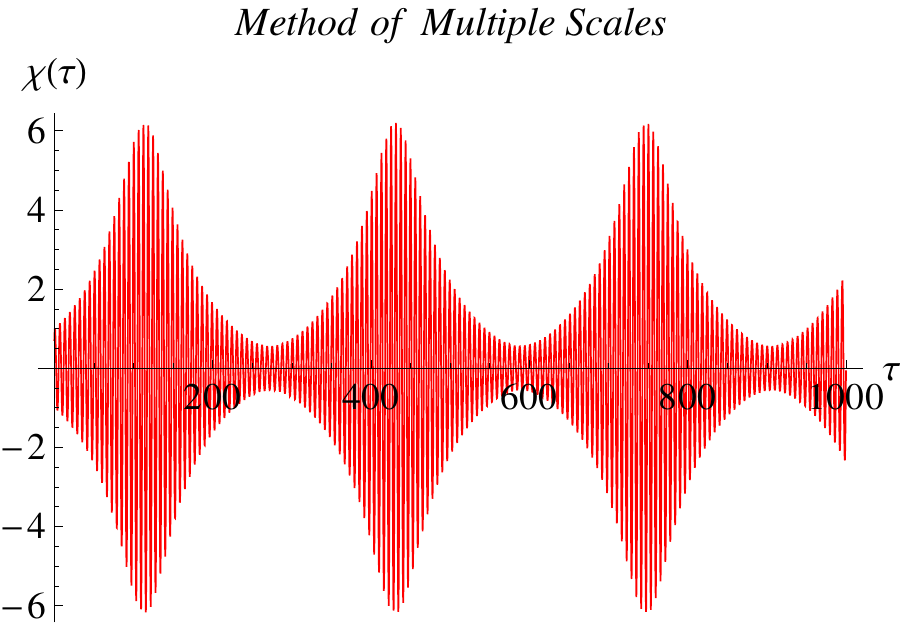}\includegraphics[width=.5\textwidth]{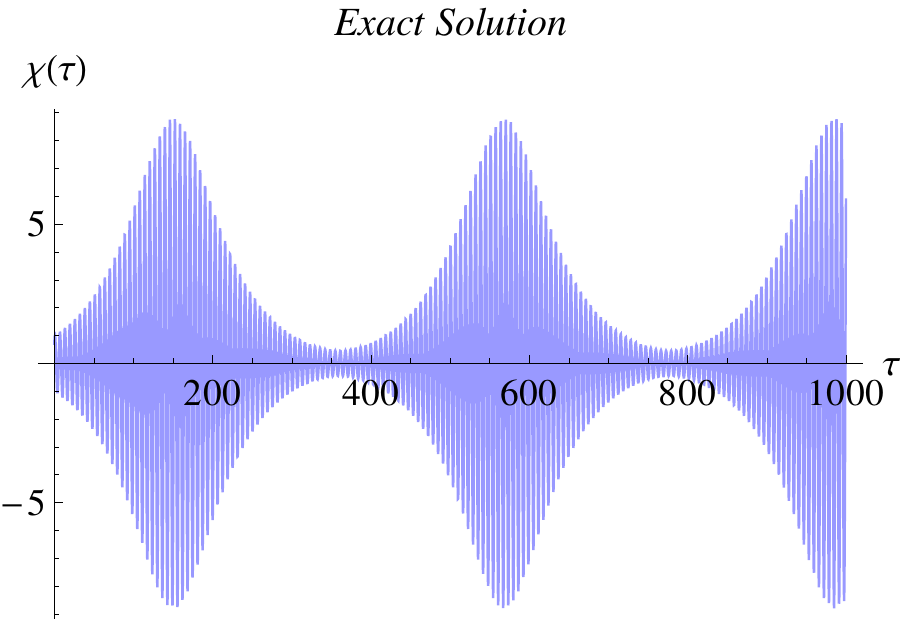}
\caption{Solutions to the relativistic Mathieu equation (\ref{RelMathieuRearranged}) through (Left) the approximation of the method of multiple scales
(\ref{MultipleScalesEqs}), 
and (Right) the exact
numerical solution, with $\hat q = 1.2,\, \hat v = 4,\, \hat \sigma = 0.3$, and $\epsilon = 0.03$. Notice that the exact and approximate solutions
have the same qualitative beating pattern; however, the amplitude and frequency of the beats are slightly different.}
\label{fig:ExactAndExpansion}
\end{figure}

\begin{figure}[t]
\centering\includegraphics[width=.6\textwidth]{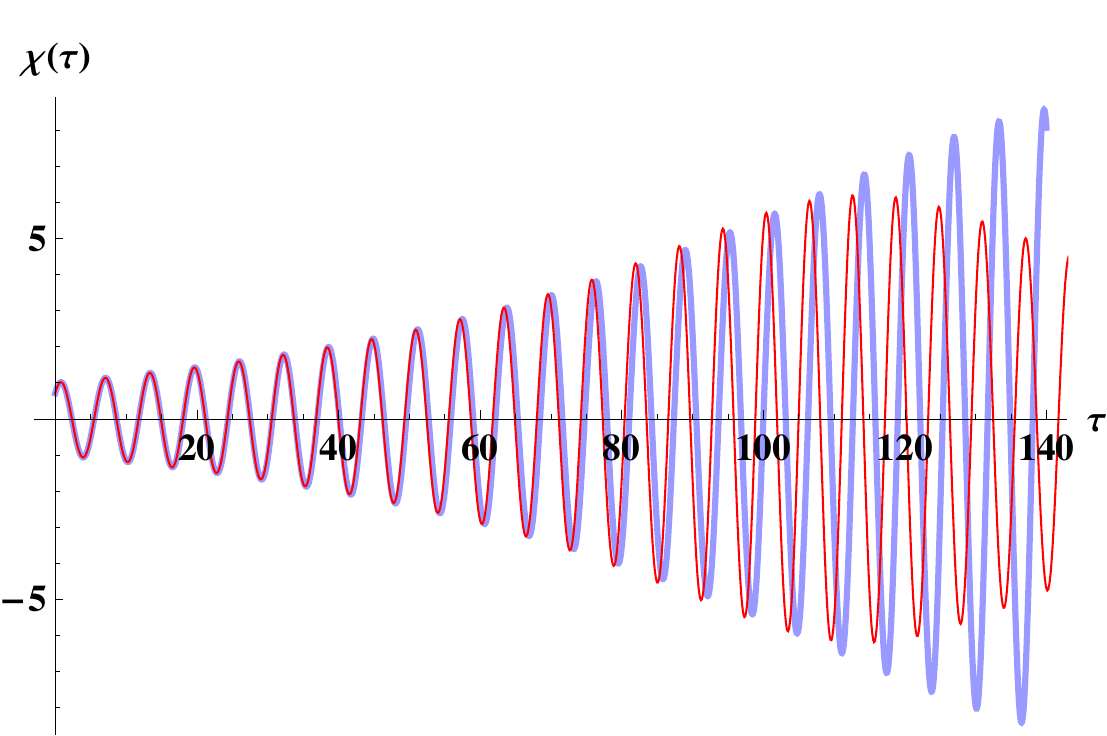}
\caption{Solutions to the relativistic Mathieu equation (\ref{RelMathieuRearranged}) through (blue, thick) the approximation of the method of multiple scales
(\ref{MultipleScalesEqs}), and (red, thin) the exact numerical solution, with the same parameters as in the previous figure. The two solutions
agree up to some time, after which they begin to diverge. Nevertheless, as we see in the previous figure, the two solutions have the same
qualitative behavior. The time over which they agree can be made longer by making $\epsilon$ smaller.}
\label{fig:ExactCompare}
\end{figure}

Solutions to (\ref{PertMathieu}) can be found perturbatively\footnote{See also \cite{Chemissany:2011nq}, which studied parametric
resonance of a similar equation using different techniques.}
using the method of multiple scales (see \cite{Nonlinear}). In particular,
to leading order the solution to (\ref{PertMathieu}) is
\be
\chi(\tau) = a(\epsilon \tau) \cos\left( A^{1/2} \tau + \beta(\epsilon \tau)\right) + {\mathcal O}(\epsilon)\, , \label{MultipleScalesAnsatz}
\ee
where the amplitude and phase $a, \beta$ are functions of the ``slow time" $T_1 = \epsilon \tau$.
In the vicinity of the first resonance band $A\sim 1$ of the non-relativistic Mathieu equation, the amplitude and phase are determined
by the equations:
\be
\frac{da}{dT_1} &=& \frac{\hat q\ a}{2 A^{1/2}} \sin \psi\, ; \nn \\
\frac{d\psi}{dT_1} &=& 2 \sigma + \frac{\hat q}{A^{1/2}} \cos \psi - \frac{3}{2} \frac{A a^2}{{\hat v}^2}\, , \label{MultipleScalesEqs}
\ee
where we made the redefinition $\psi = 2 \sigma \epsilon t - 2\beta$ and we introduced a ``de-tuning parameter" $\sigma$ defined
as $A^2 = (1-\epsilon \sigma)^2$ with $\sigma \sim {\mathcal O}(1)$.
Solutions to (\ref{PertMathieu}) through the method of multiple scales (\ref{MultipleScalesEqs})
give good qualitative agreement with soutions to the exact relativistic Mathieu equation (\ref{RelMathieuRearranged}),
as can be seen in Figure \ref{fig:ExactAndExpansion}.
We see there a characteristic beating pattern, similar to that found in the forced relativistic oscillator; however, the amplitude
and frequency of the beating is slightly different between the analytic method of multiple scales and the numerical solution. 
For our purposes in this paper, this qualitative agreement is sufficient to illustrate
the absence of unbounded resonance when the non-linear terms of (\ref{RelMathieuRearranged}) are included.
Solutions to (\ref{PertMathieu}) and (\ref{RelMathieuRearranged}) do agree at early times, as seen in Figure \ref{fig:ExactCompare};
for smaller values of $\epsilon$, the solutions can be made to agree for longer periods of time.

\begin{figure}[h]
\centering\includegraphics[width=.6\textwidth]{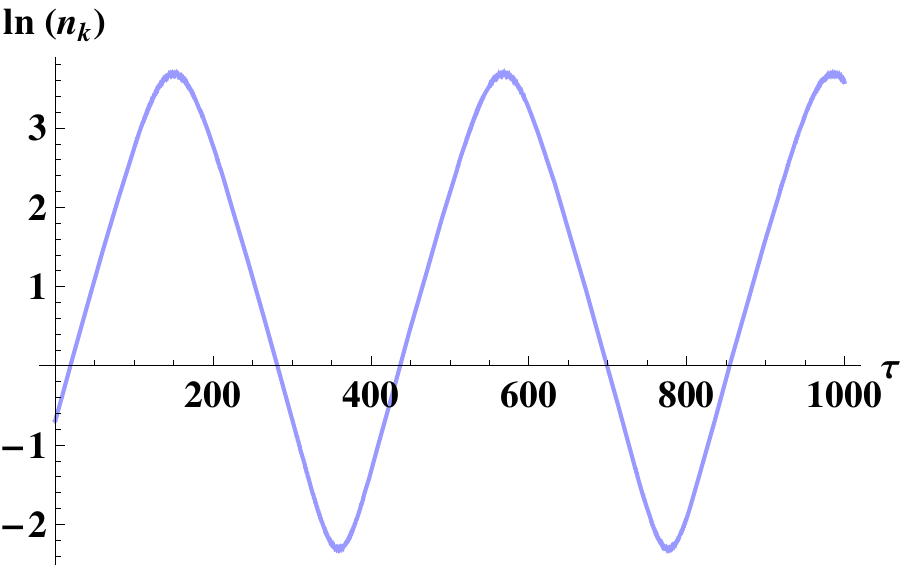}
\caption{The number density $n_k$ for the relativistic oscillator for the exact solution shown in Figure \ref{fig:ExactAndExpansion} shows that
the number density grows and falls with the amplitude $\chi$ of the reheaton field.}
\label{fig:NumberDensity}
\end{figure}

Since we are ultimately interested in the growth of particles by parametric resonance, we should also examine the behavior of the number
density of particles $n_\chi$. The number density is defined to be the energy density of a mode $\rho_\chi$ divided by the energy of that mode $\omega$;
for our relativistic oscillator, the number density is thus proportional to
\be
n_\chi = \frac{v^2}{\omega(t)} \left(\frac{1}{\sqrt{1-\chi'^2/v^2}}-1\right) + \frac{1}{2} \omega(t) \chi^2
\label{eq:NumberDensity}
\ee
where $\omega(t)^2 \equiv  A - 2 q \cos 2\tau$ is the energy of a mode. The number density for the exact numerical solution
to (\ref{RelMathieuRearranged}) shown in Figure \ref{fig:ExactAndExpansion} is shown in Figure \ref{fig:NumberDensity}. We see
clearly that the number density tracks the amplitude of $\chi$, as expected.
Note that in some models of preheating the amplitude of the reheaton field does not necessarily indicate growth of particle number \cite{preheatingLong}.
However, this failure only occurs when the amplitude of the time-dependent frequency $\omega(t)$ is changing in time, overwhelming the
time-dependence of the amplitude $\chi(t)$. For example, for the model shown in \cite{preheatingLong}, after the end of preheating the amplitude
$\chi(t)$ increases slightly while the amplitude of the frequency $\omega(t)$ also decreases slightly; the two effects balance each other out, resulting
in no particle production. This is clearly not the case here - the amplitude of $\omega(t)$ in this section is fixed, as we are considering the quantities
$A, q$ to be fixed in time. As we will show in the next section, even when we allow time-dependence for the background values of the inflaton (so that
the effective $A,q$ become time-dependent), the parametric growth of the energy density of the reheaton field is still suppressed when the non-linear
terms become important, so this particular caveat does not concern us here.

\begin{figure}[t]
\centering\includegraphics[width=\textwidth]{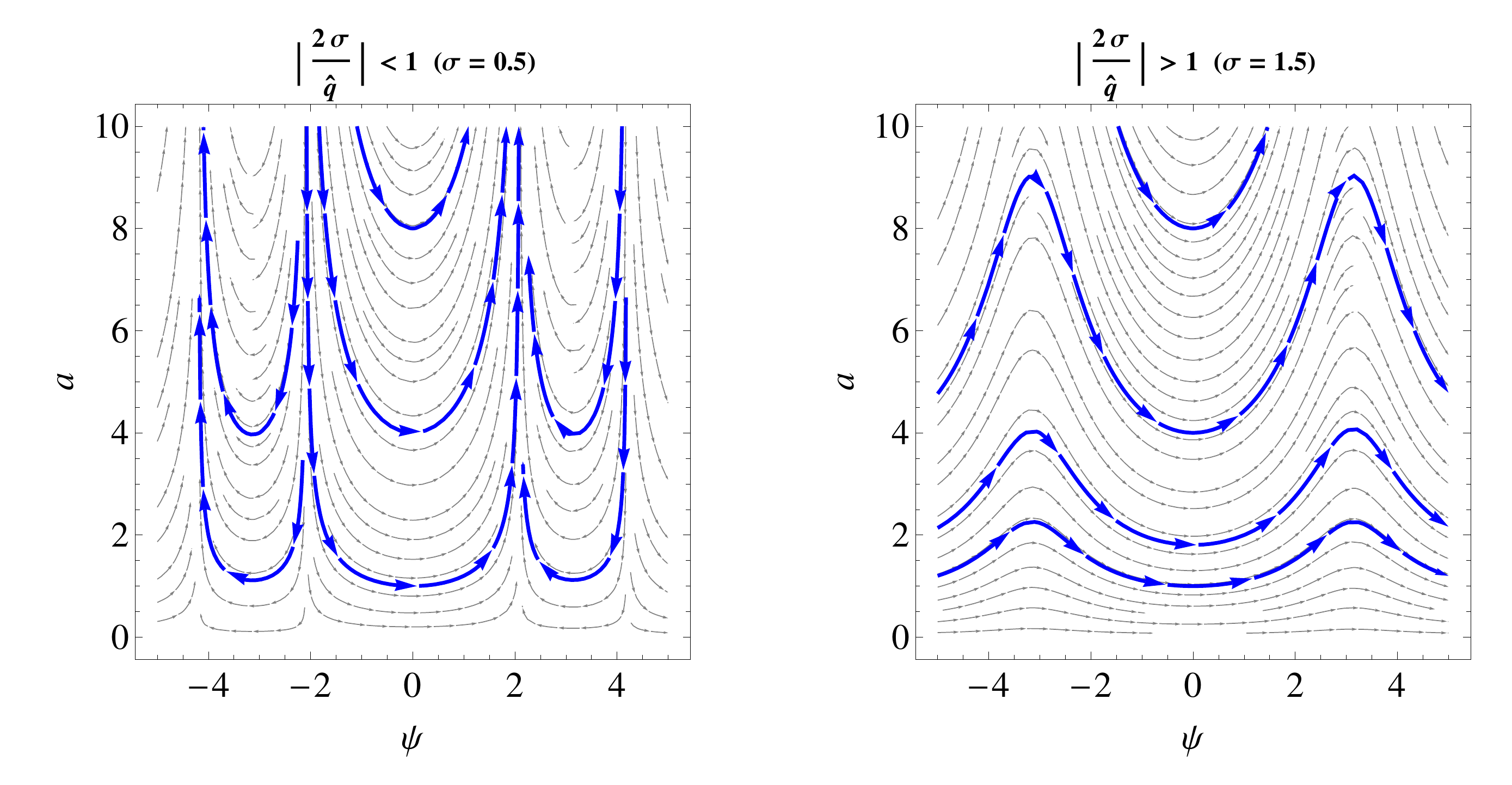}
\caption{Phase space plot of the equations (\ref{MultipleScalesEqs_NonRel}), describing the non-relativistic Mathieu equation (\ref{NonRelMathieu}),
for oscillation frequencies (Left) $|\sigma|< \hat q/2$ inside the resonance band, and (Right) $|\sigma|> \hat q/2$ outside the resonance band. 
Notice the presence of growing mode 
solutions in the resonance band (left-hand plot), indicating the presence of parametric resonance, 
while outside the resonance band solutions are bounded. The bold (blue) trajectories are some sample trajectories
to illustrate the typical behavior of solutions in phase space.}
\label{fig:NonRelStreamPlot}
\end{figure}

As a basis for comparison, let's investigate the nature of solutions in the absence
of the non-linear term from the relativistic correction, $\hat v \rightarrow \infty$.
The equation of motion (\ref{PertMathieu}) then becomes the well-known non-relativistic Mathieu equation:
\be
\chi'' + (A-2 \hat q \epsilon \cos 2\tau) \chi = 0\, . \label{NonRelMathieu}
\ee
The equations governing the amplitude and phase of (\ref{MultipleScalesAnsatz}) are then a special case of (\ref{MultipleScalesEqs}):
\be
\frac{da}{dT_1} &=& \frac{\hat q\ a}{2 A^{1/2}} \sin \psi\, ; \nn \\
\frac{d\psi}{dT_1} &=& 2 \sigma + \frac{\hat q}{A^{1/2}} \cos \psi \, . \label{MultipleScalesEqs_NonRel}
\ee

\begin{figure}[t]
\centering\includegraphics[width=\textwidth]{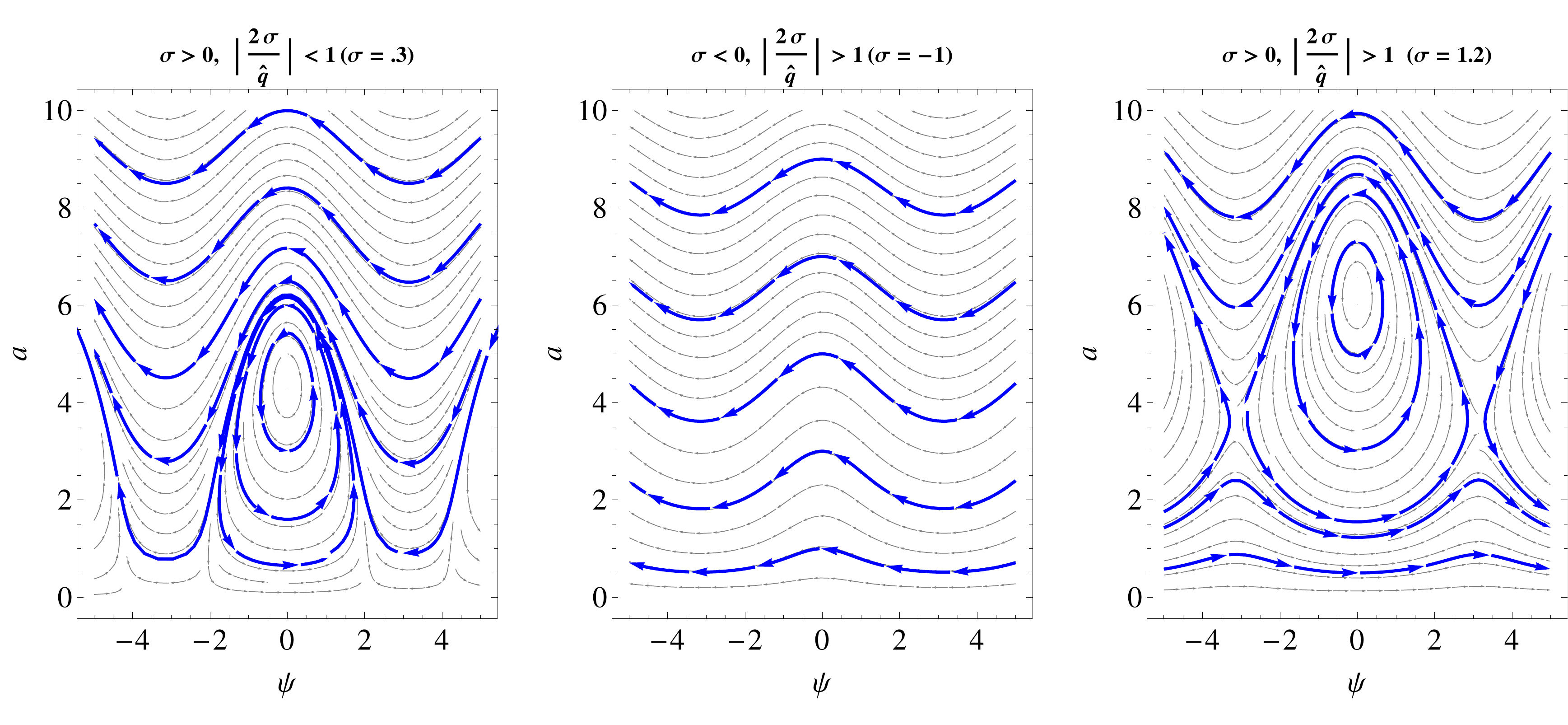}
\caption{Phase space plot of the equations (\ref{MultipleScalesEqs}), describing the relativistic correction to the Mathieu equation (\ref{PertMathieu}).
There are three regimes of interest: (Left) when the oscillation frequency is inside the traditional ``resonance band" of the non-relativistic Mathieu equation 
$|2 \sigma/\hat q| < 1$, (Center) when the oscillation frequency is below the ``resonance band", and (Right) when the oscillation frequency is above the
``resonance band". Comparing the left-hand plot to the left-hand plot of Figure \ref{fig:NonRelStreamPlot}, we see that growing mode solutions in
the non-relativistic case are now bounded when relativistic corrections are included, indicating that parametric
resonance has been suppressed. Outside of the ``resonance band", solutions are still bounded,
as seen in the center and right-hand plots. Interestingly, closed orbits develop in the relativistic case when the oscillation frequency is above the \
``resonance band", in contrast to the non-relativistic behavior.
}
\label{fig:RelUndampedStreamPlots}
\end{figure}

In this limit, there are two types of solutions:
a runaway solution with $\psi \rightarrow |\cos^{-1} \frac{2\sigma}{\hat q}|$, $a \rightarrow \infty$, and a decaying solution
$\psi \rightarrow |\cos^{-1} \frac{2\sigma}{\hat q}|$, $a \rightarrow 0$, representing the two independent growing and decaying solutions
of (\ref{NonRelMathieu}).
A phase space plot of non-relativistic solutions of (\ref{MultipleScalesEqs}) shown in Figure \ref{fig:NonRelStreamPlot} demonstrates this behavior.
Note that growing mode solutions are only possible for $|\sigma| < \hat q/2$, demonstrating the finite width of the resonance
band, while for $|\sigma| > \hat q/2$ the amplitude $a(T_1)$ is bounded.

Let us now examine the effect
of the relativistic correction to (\ref{MultipleScalesEqs}).
It is straightforward to see that there are 3 fixed points of (\ref{MultipleScalesEqs}):
\begin{enumerate}
\item[(i)] Stable fixed point $a_i = \frac{\hat v}{A^{1/2}} \sqrt{\frac{2}{3} (2\sigma + \frac{\hat q}{A^{1/2}})}$, $\psi_i = 2 n \pi$, $n \in {\mathbb N}$.
\item[(ii)] Saddle point $a_{ii} = \frac{\hat v}{A^{1/2}} \sqrt{\frac{2}{3} (2\sigma - \frac{\hat q}{A^{1/2}})}$, $\psi_{ii} = (2 n+1) \pi$, $n \in {\mathbb N}$.
\item[(iii)] Saddle point $a_{iii} = 0$, $\psi_{iii} = \cos^{-1} \left(\frac{2\sigma A^{1/2}}{\hat q}\right)$.
\end{enumerate}
Depending on the values of the parameters $\sigma, \hat q$, some or none of these fixed points will be present. The
behavior of solutions depends on the existence and types of fixed points present, so we will now discuss this in some detail.
When $2 |\sigma|/\hat q < 1$, the non-relativistic Mathieu equation (\ref{NonRelMathieu}) would have exponentially
growing solutions. The relativistic Mathieu equation, however, has fixed points of type (i) and (iii) above. Instead
of growing indefinitely, small initial amplitudes orbit the stable fixed point (i), as shown in the left-hand 
plot of Figure \ref{fig:RelUndampedStreamPlots}. These trajectories 
have a similar behavior as with the forced relativistic oscillator - after an initial period of growth, the frequency of oscillation
shifts so that $\chi(\tau)$ and the external force are no longer in phase, limiting the amount of growth.
This is the origin of the ``beating" pattern seen in the exact and multiple scales solutions shown in Figure \ref{fig:ExactAndExpansion},
and has the same physical origin as the beating as seen in the forced relativistic oscillator in Figure \ref{fig:ForcedRelOscillator}.
This qualitative understanding of the origin of the beating is justification for considering
the multiple scales approximation to solutions to the full relativistic Mathieu equation (\ref{RelMathieuRearranged}):
even though the numerical solutions to (\ref{MultipleScalesEqs}) and (\ref{RelMathieuRearranged}) are not exactly the same,
they have the same qualitative behavior, so we expect that solutions to the full relativistic Mathieu equation (\ref{RelMathieuRearranged})
are governed by phase space plots very similar to those in Figure \ref{fig:RelUndampedStreamPlots}.
We also see from Figure \ref{fig:RelUndampedStreamPlots} that if the amplitude is large enough, 
solutions change in nature from closed to open trajectories in the $(a,\psi)$ phase plane.
Regardless of initial conditions, however, we see that the amplitude is bounded and does not grow indefinitely
as in the non-relativistic case.

Continuing to other values of the parameters, when $\sigma < 0$ and $2 |\sigma|/\hat q > 1$ the behavior
is similar to the non-relativistic Mathieu equation - there are no fixed points, and solutions are oscillatory with bounded amplitude; see the center
plot of Figure \ref{fig:RelUndampedStreamPlots}.
Finally, for $\sigma > 0$ and $2 \sigma/\hat q > 1$, the non-relativistic Mathieu equation does not have resonance; however, solutions
of (\ref{MultipleScalesEqs}) in this parameter range have the same behavior as solutions in the relativistic ``resonance band"  $2 |\sigma|/\hat q < 1$,
with fixed points of type (i) and (ii). Solutions for $a, \psi$ form orbits in the neighborhood of the stable fixed point (i), while for
large amplitudes trajectories become open with bounded amplitude, as seen in the right-hand plot of Figure \ref{fig:RelUndampedStreamPlots}.

To conclude this subsection, let us summarize our findings. Based on qualitative arguments, we expect that solutions of the relativistic Mathieu
equation do not exhibit any resonance or unbounded growth, even when the corresponding Mathieu equation {\it does} have resonant growth.
We have shown this explicitly by reducing the full relativistic Mathieu equation (\ref{RelMathieuRearranged}) to the equations
(\ref{MultipleScalesEqs}) for the amplitude and phase of the solution (\ref{MultipleScalesAnsatz}).
The phase diagram of (\ref{MultipleScalesEqs}) for the values of the parameters that would normally give rise to resonance
instead contains a stable fixed point at finite amplitude, which nearby solutions orbit, as shown in Figure \ref{fig:RelUndampedStreamPlots}.
At large amplitudes, the phase diagram of (\ref{MultipleScalesEqs}) resembles that of the non-resonant Mathieu equation --- solutions
are oscillatory with bounded amplitude.

In the next section, we will discuss how the inclusion of non-zero damping in (\ref{RelMathieuRearranged}) modifies these observations.

%%%%%%%%%%%%%%%%%%%%%%%%%%%%%%%%%%%%%%%
\subsection{Damped Relativistic Mathieu Equation}
\label{sec:DampedRel}
%%%%%%%%%%%%%%%%%%%%%%%%%%%%%%%%%%%%%%%

Now let us include non-zero damping in the relativistic Mathieu equation,
\be
\chi''(\tau) + \mu \chi' \left(1-\frac{\chi'^2}{v^2}\right)^{3/2}+ (A-2 q \cos 2\tau) \left(1-\frac{\chi'^2}{v^2}\right)^{3/2} \chi(\tau) = 0\, . \label{DampedRelMathieuRearranged}
\ee
We expect similar behavior for the damped Mathieu equation above as from the previous section --- as the amplitude and speed increase, 
the damping and effective frequency vanish so that (\ref{DampedRelMathieuRearranged}) becomes the equation for a free particle without
resonance. Further, as the amplitude of $\chi$ increases, so does the frequency of oscillation of $\chi$, so that if $\chi$ does experience
resonance initially, it soon becomes out of phase with the time-dependent effective frequency and resonance quickly ceases.

We can construct semi-analytic solutions to (\ref{DampedRelMathieuRearranged}) as in the previous subsection: we let
$q = \epsilon \hat q$, $v^2 = \hat v^2 \epsilon^{-1}$, and $\mu = \epsilon \hat \mu$, where 
$\epsilon \ll 1$ and $\hat q, \hat v, \hat \mu \sim {\mathcal O}(1)$. Expanding (\ref{DampedRelMathieuRearranged}) to leading order, 
we have,
\be
\chi'' + \epsilon \hat \mu \chi' + (A-2\epsilon \hat q \cos 2\tau)\chi = \frac{3}{2} \epsilon A \frac{\chi'^2}{\hat v^2} \chi\, . \label{DampedRelMathieuPert}
\ee
Solutions to (\ref{DampedRelMathieuPert}) have the same form as in the undamped case:
\be
\chi(\tau) = a(\epsilon \tau) \cos \left(A^{1/2} \tau + \beta(\epsilon \tau)\right) + {\mathcal O}(\epsilon)\,. \label{DampedRelMathieuAnsatz}
\ee
In the vicinity of the first resonance band, the amplitude and phase of (\ref{DampedRelMathieuAnsatz}) obey the equations:
\be
\frac{d a}{dT_1} &=& \frac{\hat q\ a}{2 A^{1/2}} \sin \psi - \frac{\hat \mu}{2} a \nonumber \,; \\
\frac{d \psi}{dT_1} &=& 2 \sigma + \frac{\hat q}{A^{1/2}} \cos \psi - \frac{3}{2} \frac{A a^2}{\hat v^2} \,, \label{DampedRelMathieuEqs}
\ee
where $\psi, \sigma$ are defined as in the previous subsection.

\begin{figure}[t]
\centering\includegraphics[width=.55\textwidth]{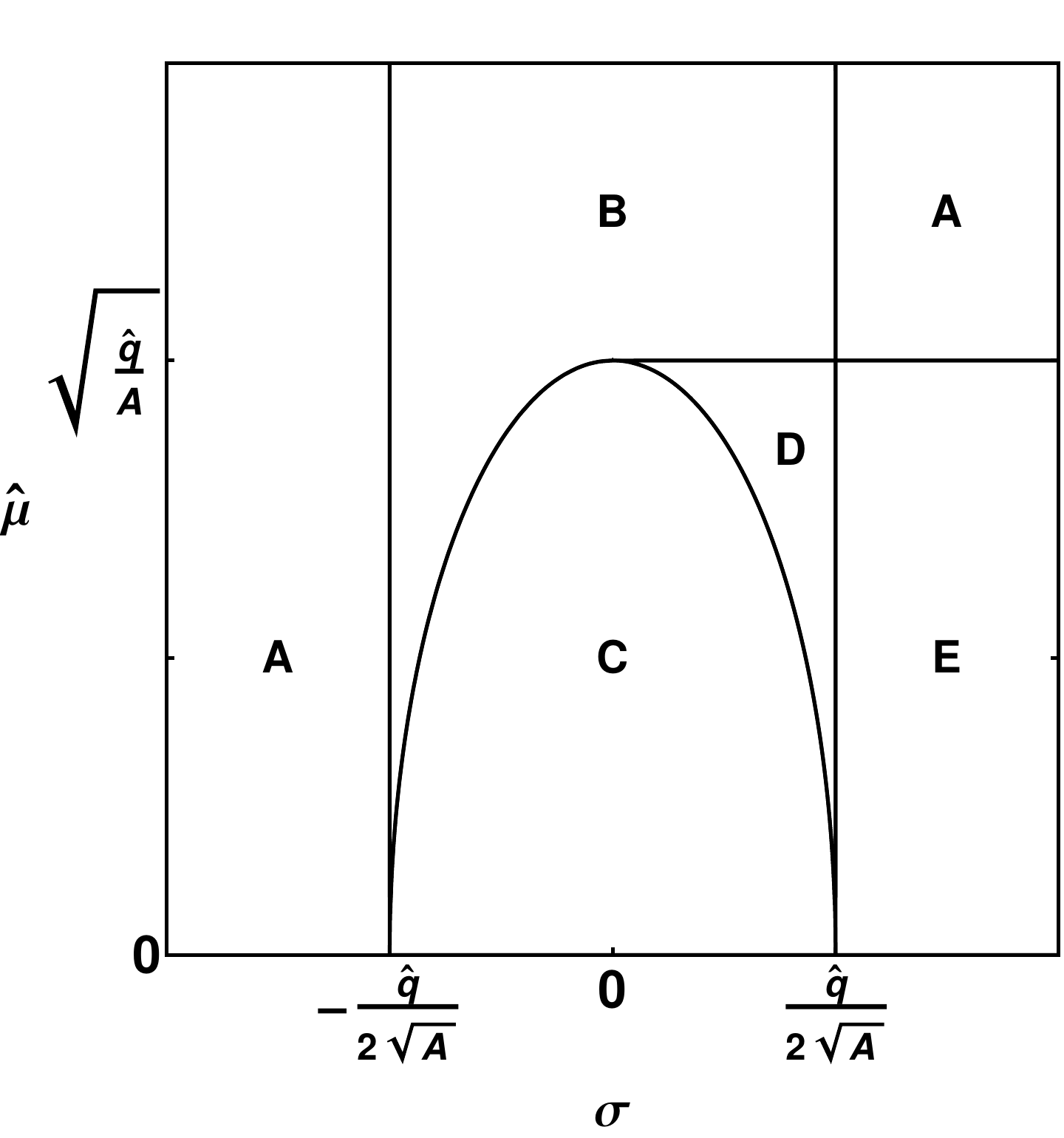}
\caption{Stability chart for the damped Mathieu equation. See text below (\ref{DampedRelMathieuEqs}) for explanation.}
\label{fig:StabilityChartDamp}
\end{figure}

The fixed points of (\ref{DampedRelMathieuEqs}) are now shifted to:
\begin{enumerate}
\item[(i)] Fixed point 
\be
a_i = \frac{\hat v}{A^{1/2}} \sqrt{\frac{2}{3} \left(2\sigma + \frac{\hat q}{A^{1/2}}\sqrt{1-\hat\mu^2 A/\hat q}\right)}, 
	\hspace{.3in} \psi_i = \left| \cos^{-1} \left(\sqrt{1-\hat \mu^2 A/\hat q}\right)\right| \nn
\ee
\item[(ii)] Fixed point
\be
a_i = \frac{\hat v}{A^{1/2}} \sqrt{\frac{2}{3} \left(2\sigma - \frac{\hat q}{A^{1/2}}\sqrt{1-\hat\mu^2 A/\hat q}\right)}, 
	\hspace{.3in} \psi_i =-\left| \cos^{-1} \left(\sqrt{1-\hat \mu^2 A/\hat q}\right)\right| \nn
\ee
\item[(iii)] Fixed point $a_{iii} = 0$, $\psi_{iii} = \cos^{-1} \left(\frac{2\sigma A^{1/2}}{\hat q}\right)$.
\end{enumerate}

The pattern of fixed points is much more complicated than in the undamped case, as seen in Figure \ref{fig:StabilityChartDamp}.
The values of the parameters $\hat \mu, \sigma$, affect the existence and stability of the fixed points.
In Region {\bf A}, no stable fixed points exist; trajectories are bounded and open, as in the middle plot of Figure \ref{fig:RelUndampedStreamPlots}.
 In Region {\bf B}, the only fixed points are those of type (iii), which are stable.
Notice that this means that in Region {\bf B}, solutions decay to zero amplitude - there are no fixed points
with non-zero amplitude.
In Region {\bf C}, fixed points of type (i) and (iii) both exist, but only the fixed points of type (i) are
stable in this region - here there are fixed points with non-zero amplitudes.
In Region {\bf D}, fixed points of type (i), (ii) and (iii) all exist simultaneously, 
and the fixed points of type (i) and (iii) are both stable in this region.
Finally, Region {\bf E}, fixed points of type (i) and (ii) exist, but only those of type (i) are stable here.

\begin{figure}[t]
\centering\includegraphics[width=.7\textwidth]{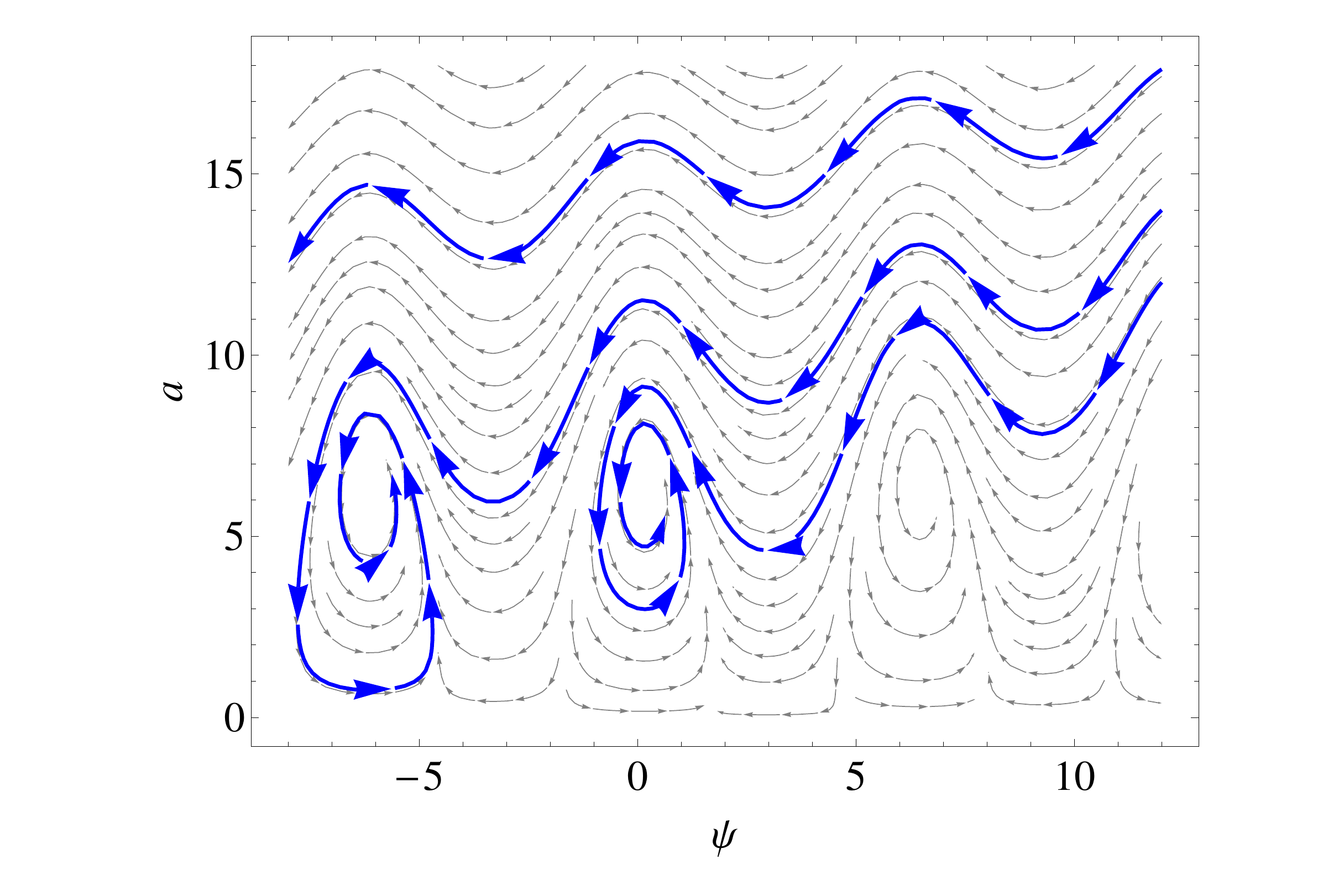}
\caption{Phase space plot of the equations (\ref{DampedRelMathieuEqs}) describing the damped relativistic Mathieu equation inside
the usual ``resonance band" $|2\sigma/\hat q| < 1$ and $\sigma = 0.50$.
In contrast to Figure \ref{fig:RelUndampedStreamPlots}, orbits are no longer closed but instead decay to the stable fixed point. Most importantly,
there is no exponential growth inside the traditional ``resonance band" of the non-relativistic Mathieu equation.}
\label{fig:DampedMathieu}
\end{figure}

Despite the more complicated pattern of fixed points, 
many of the same features persist as in the undamped case. First, there are no resonant $a \rightarrow \infty$ solutions
as there are in the non-relativistic Mathieu equation. Even with damping, the non-relativistic Mathieu equation allows for resonant
solutions. However, the presence of the relativistic correction term in (\ref{DampedRelMathieuRearranged},\ref{DampedRelMathieuPert}) affects
the phase of the oscillation at large amplitude as we can see in (\ref{DampedRelMathieuEqs}), effectively shutting down
the resonance.
The presence of damping does affect the nature of the solutions in one qualitative way --- in the presence of damping, stable fixed
points become attractors. At late times, all solutions eventually end up at the attractor solutions, as demonstrated in
Figure \ref{fig:DampedMathieu}.

Thus, again we see that the non-linearities of the relativistic corrections to (\ref{DampedRelMathieuRearranged}) prevent parametric
resonance from occurring.

%%%%%%%%%%%%%%%%%%%%%%%%%%%%%%%%%%%%%%%
%%%%%%%%%%%%%%%%%%%%%%%%%%%%%%%%%%%%%%%
\section{Implications for Preheating}
\label{sec:Preheating}
%%%%%%%%%%%%%%%%%%%%%%%%%%%%%%%%%%%%%%%
%%%%%%%%%%%%%%%%%%%%%%%%%%%%%%%%%%%%%%%
 Now let us apply our results from the previous section to preheating with a non-canonical kinetic term.
Specifically, we will numerically solve for the dynamics of the equations of motion (\ref{HEOM}-\ref{eq:ChiEOM1}), explicitly
allowing for time-dependence in the coefficients of the $\chi$ equation of motion (\ref{eq:ChiEOM1}).
As we found in the previous section, including the non-linearities from the reheaton's non-canonical kinetic
terms affects the existence of parametric resonance.

\begin{figure}[t]
\centering\includegraphics[width=\textwidth]{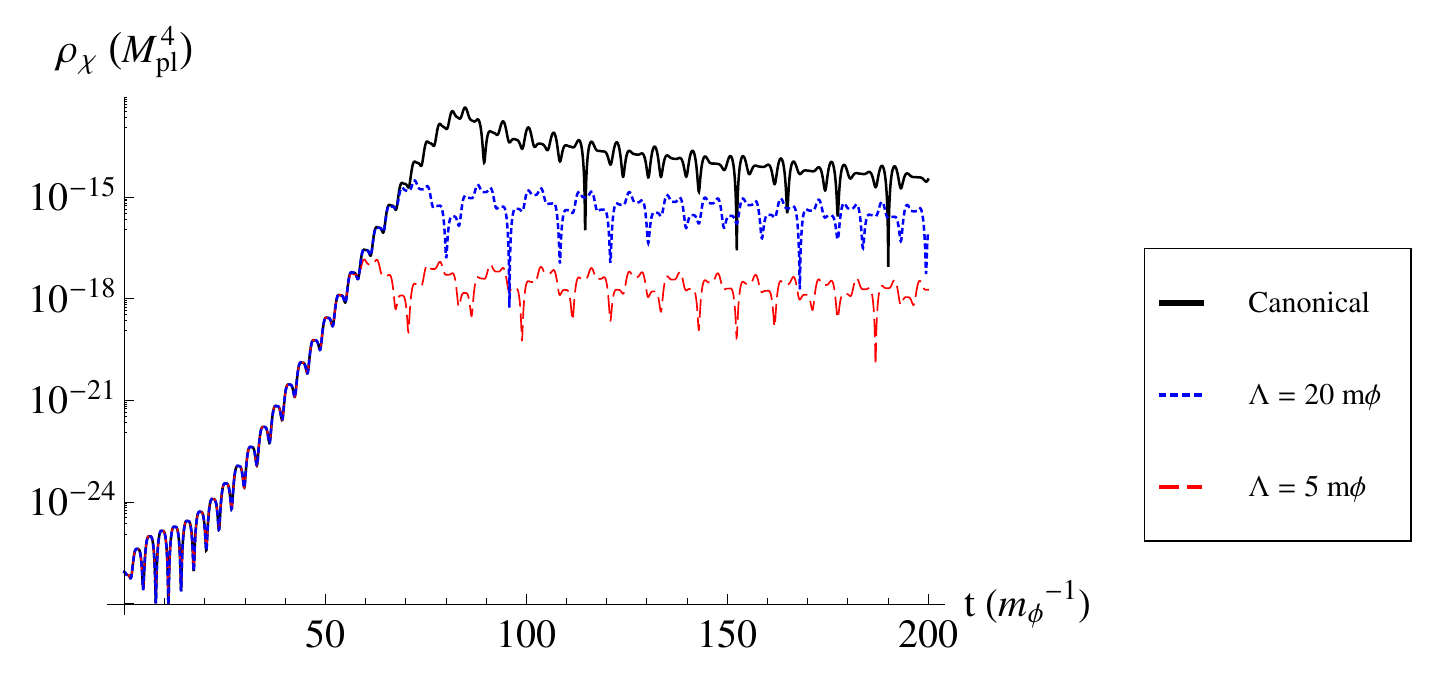}
\caption{The energy density of the reheaton field during preheating for different values of the UV scale $\Lambda$. Notice that the non-canonical
reheaton field initially behaves like a canonical field and grows through resonance, but when the non-canonical kinetic terms
become important the field falls out of resonance at an early time, suppressing the amount of energy transferred to the reheaton field.}
\label{fig:PreheatingPlots}
\end{figure}

Let us numerically investigate a specific example, with
the parameter values $g= 2\times 10^{-3}, m_\phi = 10^{-5} M_p, \Phi_0 = 0.01 M_p, m_\chi = 10^{-7} M_p$ and $\chi(0) =\frac{H}{2\pi} \sim 6\times 10^{-9} M_p$.
Since perturbations of massive scalar fields should be damped for long-wavelengths, the latter initial condition is chosen to be a concrete 
(and not arbitrary) initial condition, as would be expected for large wavelength modes of a light scalar field in a nearly 
de-Sitter background \cite{Chaotic,Wands:2007bd}.
As discussed in the introduction, the parameter $\Lambda$ corresponds to some UV degrees of freedom that have been integrated out.
For consistency of the effective theory, then, $\Lambda$ cannot be too small; in particular, it should at least be larger
than the mass of the inflation $\Lambda > m_\phi$.
Numerical solutions of (\ref{HEOM}-\ref{eq:ChiEOM1}) for the energy density $\rho_\chi$
are shown in Figure \ref{fig:PreheatingPlots}
for several different values of the parameter $\Lambda$ as well as the energy density for a reheaton with a canonical kinetic term.

A few things are worth noting here. First, the non-canonical reheaton field
behaves much like a canonical reheaton field for the initial stages of preheating. However, as the amplitude of $\chi$ increases
the non-canonical reheaton field eventually falls out of resonance with the oscillating effective frequency, and preheating terminates
at an earlier time and lower energy density.
Thus, a smaller number density of $\chi$ particles is produced by preheating, and we thus expect similar reductions in possible violations
of adiabaticity conditions.
Second, as $\Lambda$ increases, the final energy density in the reheaton field approaches that of a canonical reheaton field,
as expected.
Note that the parameters chosen above imply $q_0 = 1, A_0 = 2.0$,
and $\mu = 1200$ in terms of the parameters of (\ref{eq:ChiEOM3},\ref{MathieuParameters}).
Thus, while the analysis of the previous sections is strictly only valid for small $q$, the lessons learned are valid for
a much wider range of parameters.

Finally, we note that while our analysis has so far been restricted to the DBI Lagrangian (\ref{DBI}), many different types of non-linear
self-interactions may have similar behavior during preheating.
In particular, the quartic self-interaction:
\be
\t p(X_\chi,\chi) = \frac{1}{2} (\partial \chi)^2 - \frac{1}{2} m_\chi^2 \chi^2 - \frac{1}{4} \lambda \chi^4\,,
\ee
gives rise to a cubic-order term in the equation of motion for $\chi$ similar to the term on the right hand side of (\ref{PertMathieu}).
The equation of motion then takes the form of the parametrically excited Duffing equation, which exhibits similar behavior as the DBI
Lagrangian studied here \cite{Nonlinear}.
Similarly, keeping one of the leading non-renormalizable operators in the expansion (\ref{eq:AllNCLagrangian}):
\be
\t p(X_\chi,\chi) =  \frac{1}{2} (\partial \chi)^2 - \frac{1}{2} m_\chi^2 \chi^2 + c_8 \frac{(\partial \chi)^4}{\Lambda^4}\,,
\ee
will also lead to an equation of motion for $\chi$ similar to (\ref{PertMathieu}). Thus, while our analysis has been restricted to the
DBI Lagrangian, the lessons learned are more general.

%%%%%%%%%%%%%%%%%%%%%%%%%%%%%%%%%%%%%%%
%%%%%%%%%%%%%%%%%%%%%%%%%%%%%%%%%%%%%%%
\section{Conclusion}
\label{sec:Conclusion}
%%%%%%%%%%%%%%%%%%%%%%%%%%%%%%%%%%%%%%%
%%%%%%%%%%%%%%%%%%%%%%%%%%%%%%%%%%%%%%%

We have investigated preheating when the reheaton field has non-canonical kinetic terms of the DBI type.
The equation of motion for long-wavelength modes is the same as for a damped relativistic oscillator with
time-dependent effective frequency --- the relativistic Mathieu equation. In contrast to the usual Mathieu equation,
the relativistic Mathieu equation does not exhibit unbounded parametric resonance, so that growth in the reheaton field
cannot be arbitrarily large. The primary reason for the absence
of unbounded resonance is that the relativistic oscillator is {\it anharmonic} --- the frequency of oscillation depends on the amplitude.
When the amplitude is initially small, the relativistic Mathieu equation behaves like the non-relativistic Mathieu equation
and solutions grow due to resonance in the usual resonance bands.
However, as the amplitude grows the frequency changes, and the reheaton is soon out of phase
with the oscillating effective frequency.
This effect limits the growth of the reheaton field with non-canonical kinetic terms. We studied this in some
detail by finding perturbative solutions to the full relativistic Mathieu equation, and verified that this effect limits the growth
of the reheaton field during preheating.

We made a number of simplifications in order to make progress on this question. Most significant, we considered only
long-wavelength modes of the reheaton field. In non-linear systems such as the one studied in this paper, different
Fourier modes with finite wavelength do not decouple can interact in complex and interesting ways --- a full analysis of the $k\neq 0$
case requires a complex lattice simulation.
Further, we did not include backreaction of the reheaton field on the expanding background and inflaton field, or
rescattering of fluctuations, all of which are important effects \cite{preheatingLong}. Instead, our focus was only on the non-linear terms
introduced by non-canonical kinetic terms in the long-wavelength approximation. A firmer understanding of the effects of these terms on parametric resonance
in this simple case will help our intuition for when we eventually can relax these simplifications.

%%%%%%%%%%%%%%%%%%%%%%%%%%%%%%%%%%%%%%%
%%%%%%%%%%%%%%%%%%%%%%%%%%%%%%%%%%%%%%%
\subsection*{Acknowledgments} 
This research was supported by a 
Pacific Lutheran University Division of Natural Sciences research grant.
%%%%%%%%%%%%%%%%%%%%%%%%%%%%%%%%%%%%%%%
%%%%%%%%%%%%%%%%%%%%%%%%%%%%%%%%%%%%%%%

\bibliography{refs}

\bibliographystyle{utphysmodb}

\end{document}